\documentclass[12pt]{article}
\usepackage{graphicx,graphics}
\title{Distillation of Bell states in open systems.}
\author{E. Isasi and  D. Mundarain
\\
\\
Departamento de F\'{\i}sica, Secci\'{o}n de Fen\'{o}menos \'{O}pticos, \\
Universidad Sim\'{o}n Bol\'{\i}var,\\
Apartado Postal 89000, Caracas 1080A, Venezuela }

\begin{document}
\maketitle
\begin{abstract}
In this work we review the entire classification   of $2\times2$ distillable states for  protocols with  a finite numbers of copies. We show a distillation protocol that allows to distill Bell states with non zero probability at any time for an initial singlet in vacuum. It is shown that the same protocol used in non zero thermal baths yields  a considerable recovering of entanglement.
\end{abstract}

\section{Introduction}
Maximal entangled states are the basic resource for the encoding and communication on quantum information \cite{EinAPR1935,HorRHH2007,BenCBC1993}.  In open systems entanglement degrades  and it is necessary to design strategies that   preserve  or  partially recover  the entanglement. Error Correction Codes\cite{BenDSW1996}, Filtering and Distillation of entanglement \cite{VerFDD2001,BenBPS1996} and the use of Decoherence Free Sub-Spaces \cite{Lidar}, are some of such strategies.

In a typical protocol for distillation of entanglement there are two observers, Alice and Bob  and  two o more copies of bipartite entangled system.  Each observer has access only to one part of the system on which he or she may  applicate local operations. These operations can be independent ( filtering) or collectives ( proper distillation). In a proper distillation protocol at the end of the process, after the collective operations have been completed, the observers execute a set of local measurements on some copies ( the ancillas ) leaving one copy ( the source) untouched. By the effect of measurements the ancillas  collapse to separable states, transferring its entanglement to the undisturbed copy . 

The most used protocols, i.e.  Bennett's protocol \cite{BenBPS1996} and 
Deutsch's protocol \cite{DeuEJMPS1996}, use Bell diagonal states as the resource from which  to extract entanglement. In fact, Bennett's protocol starts with a Werner state obtained by applying a random bilateral rotation to an arbitrary state. Maximal entangled states are obtained only after an infinite distillation steps.

In \cite{VersDD2002}, Verstraete \textit{et. al.} show that all the mixed states of a 2-qubits system are classified in two equivalence classes. The first class is formed by the  states which are equivalent to a Bell-diagonal state under local $SL(2,C)$ transformations. The second class is formed by states  which are represented by non-diagonal matrices (in the Bell basis) parametrized by four real parameters. It comprises four sub-sets depending on the values of the parameters.

In \cite{ChenL2001}, the authors advocate  the use of non quasi-separable states (see section 2)  to distill states with maximal entanglement in a finite number of distillation steps. In this work we first review the structure of quasiseparable states for $2\times2$ systems and  show that all non-quasiseparable states are non diagonal and belong to one of the four sub-sets mentioned before. 

This study  allows to define a finite round protocol for distilling, at any time, a maximal entangled state in a two qubits system at zero temperature. Starting with singlet, it is shown that the system evolves inside the set of non-quasiseparable states, allowing the application of the protocol at arbitrary times. For thermal baths at non zero temperature the states of the systems are always quasi-separable, so it is not possible to distill a Bell state. Nevertheless under certain conditions it is possible to increase the entanglement applying the same protocol used for vacuum.

\section{Quasi separable states.}
Consider a mixed state $\rho=\sum p_i|\Psi_i\rangle\langle\Psi_i|$, with non vanishing probabilities $p_i$ and the transformations that change the probabilities  without adding or
suppressing any pure state in the mixture:
\begin{equation}
\rho=\sum p_i|\Psi_i\rangle\langle\Psi_i| \rightarrow \quad  \rho'=\sum p'_i|\Psi_i\rangle\langle\Psi_i|
\end{equation}
The new probabilities $p'_i$ should be also non vanishing, and we will refer to $\rho'$ as a
new state of $ \rho$. By definition a  state $\rho$ is called  quasi-separable if at least one of
its new states is separable.

If a given final state  is obtained through a distillation process using $n$ copies of the original state, then the same final state may be obtained with  different probability from any of the new states of $\rho$. This implies in particular that  is not possible to distill a maximal entangled state using a finite number of copies if the source or the ancilla are in quasiseparable states. In order to distill Bell states, is useful
to determine which states of the $2\times2$ system are not quasiseparable.

On the other hand, if one can distill a Bell state from a given state obviously one can distill the same Bell state from  any other state obtained by local operations and classical communication.  Now we consider the states of the system inside the equivalent classes defined by Verstraete {\it et al} in \cite{VersDD2002}. The first class is composed of Bell-diagonal states, they can be written as:
\begin{equation}
 \rho_{bd} = \sum_{i=1}^{4} P_i |\psi_i\rangle\langle\psi_i|\;,
\end{equation}
with $|\psi_i\rangle$ the four Bell states. For any mixed Bell diagonal state one can observe that it is possible to find a new  state for which all the probabilities are the same, this new state is separable without mattering the rank of the original matrix. So, excluding Bell states, all Bell diagonal states are quasi-separable and are not useful in order to distill a Bell State.

Following the classification, the second class is formed by non diagonalizable states represented by matrices of the form:
\begin{equation}
\rho=\frac{1}{2}\left(\begin{array}{cccc}
b+c & 0 & 0 & 0\\
0 & a-b & d & 0\\
0 & d & a-c & 0\\
0 & 0 & 0 & 0
 \end{array}\right)\,.
\end{equation}
Here the entries $\{a,b,c,d\}$  satisfy one and only one of the following four conditions:
\begin{enumerate}
 \item $b=c=\frac{a}{2}$
 \item $d=c=0$ and $a=b$
 \item $d=b=0$ and $a=c$
 \item $d=0$ and $a=b=c$
\end{enumerate}

The cases 2,3 and 4 correspond to  separable states from which it is not possible to distill any entanglement.

In the first case one  is compelled to consider two kind of states. For a non conditioned $d$ the  rank of the matrix is 3 and the state takes the form:

\begin{equation}\label{uto1}
\rho=\frac{1}{2}\left(\begin{array}{cccc}
a & 0 & 0 & 0\\
0 & a/2 & d & 0\\
0 & d & a/2 & 0\\
0 & 0 & 0 & 0\\
\end{array}\right)
\end{equation}
Requiring normalization one puts  $a=1$ and  $\rho$  can be decomposed as follows,
\begin{equation}
\rho=\frac{1}{2}|++\rangle \langle++| +\frac{1}{2}(\frac{1}{2}+d)|\Phi^+\rangle \langle \Phi^+| +\frac{1}{2}(\frac{1}{2}-d)|\Phi^-\rangle \langle \Phi^-| \
\end{equation}
with
\begin{equation}
 |\Phi^{\pm}\rangle = \frac{1}{\sqrt{2}} \left( |+-\rangle \pm  |-+\rangle \right)
\end{equation}
The state $\rho'$ defined by
\begin{equation}
\rho'=\frac{1}{2}|++\rangle \langle++| +\frac{1}{4}|\Phi^+\rangle \langle \Phi^+| +\frac{1}{4}|\Phi^-\rangle \langle \Phi^-| \
\end{equation}
is a new separable state of $\rho$, and so all the states of form (\ref{uto1}) with unconditioned $d$ are quasiseparable.

Consider now the states with matrices of rank 2  defined by setting the coefficients  $d=b=c=a/2= 1/2$. The density matrix in this case is :
\begin{equation}
\rho=\frac{1}{2}|++\rangle \langle++| +\frac{1}{2}|\Phi^+\rangle \langle \Phi^+| 
\end{equation}
any new state of this $\rho$ can be written as 
\begin{equation}\label{ito2}
\rho'= (1-P_1)|++\rangle \langle++| +P_1|\Phi^+\rangle \langle \Phi^+| 
\end{equation}
The concurrence of this state is $C=P_1$. Since by definition  $P_1\in(0,1)$, it is clear that the concurrence is different from zero for any choice of $P_1$. The main concussion is that these states are not quasiseparable.

\section{Distillation protocol with non quasi-separable states}

In this section we  review briefly the distillation protocol introduced by Cheng {\it et. al.} in Ref. \cite{ChenL2001}. Suppose that one has a set of $2\times2$ systems which are in the state  (\ref{ito2}).  Use one half of the systems as a source and the other half as the ancilla. Alice   execute a unilateral  NOT operation on the  source  particles, then  state of the source  becomes : 
\begin{equation}
\rho_s= (1-P_1)|-+\rangle \langle-+| +P_1|\Psi^+\rangle \langle \Psi^+| 
\end{equation}
The ancilla remains in the original state.

Now Alice and Bob execute unilateral c-not operation on their respective particles  regarding the source as control and the ancilla as target.  Finally Alice and Bob make a measurement in the computational basis of the ancilla. If they get the $+ -$ result the ancilla collapses to $|+-\rangle$ and the source to $ |\Psi^{+}\rangle $. The success probability in this case is $P_s= P_1^2/2$.

One can increase the success probability allowing the use of the $-+$ as a valid result. In that case the final reduced density matrix of the source is not more the Bell state but an state with same structure than the original one: 
\begin{equation}
\rho_s= \frac{(1-P_1)^2}{(1-P_1)^2+P_1^2}|-+\rangle \langle-+| +\frac{P_1^2}{(1-P_1)^2+P_1^2}|\Psi^+\rangle \langle \Psi^+| 
\end{equation}
The concurrence of this state is $C =\frac{P_1^2}{(1-P_1)^2+P_1^2}$ and the
probability of success is $P_s =(1-P_1)^2+P_1^2$. In Figure (1) we plot the
concurrence of the distilled state and the  concurrence of the non-distilled state. As one
can see the protocol that include both results increases the entanglement only
when $P_1> 1/2$.
\begin{figure}\label{fig11}
\centerline{\includegraphics[scale=1.0]{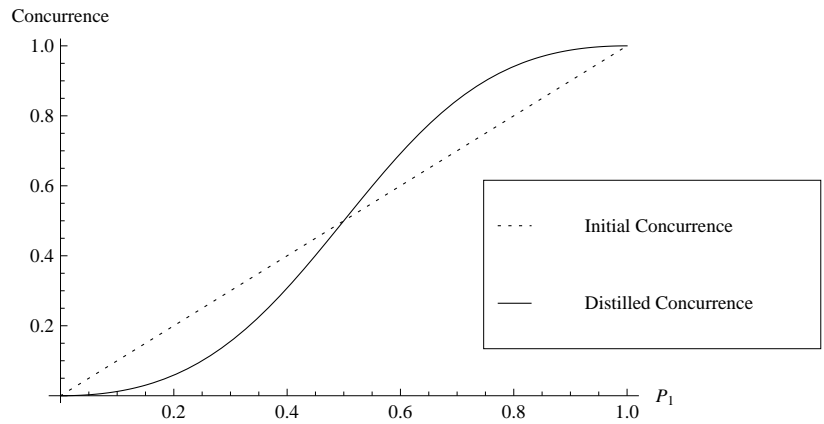}}
\caption{Initial and distilled concurrence when one takes $+-$ and $-+$ result}
\end{figure}

In Figure (2) we compare the probability when one takes only the result $+-$ with that when one takes the result $+-$ and $-+$
\begin{figure}\label{fig2}
\centerline{\includegraphics[scale=1.0]{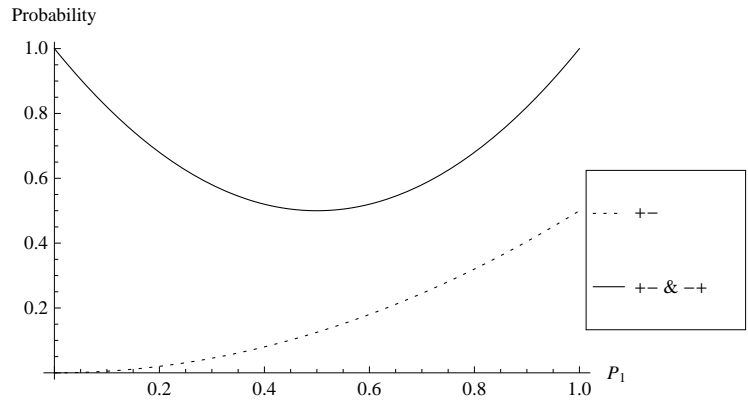}}
\caption{Probability when one take the result $+-$ and when one takes $+-$ and $-+$ results}
\end{figure}

\section{The singlet in vacuum.}
The master equation for a system composed  of two two-level particles in
vacuum is
\begin{eqnarray}
\dot{\rho} &= &\frac{\gamma}{2}\left( ( 2 \sigma_a \rho \sigma_a^{\dagger} -\sigma_a^{\dagger}\sigma_a \rho -\rho \sigma_a^{\dagger}\sigma_a ) \right.\nonumber\\
&+&\left.  ( 2 \sigma_b \rho \sigma_b^{\dagger} -\sigma_b^{\dagger}\sigma_b \rho -\rho \sigma_b^{\dagger}\sigma_b )\right)
\end{eqnarray}
where $\sigma_a= \sigma_a\otimes1$ and  $\sigma_b= 1\otimes\sigma_b$.  We
assume  that  the baths associated with each particle are independent.

If the initial state is the singlet the solution is:
\begin{equation} 
\rho (t)= (1-e^{-\gamma t})|--\rangle \langle--| +e^{-\gamma t}|\Phi^-\rangle \langle \Phi^-|
\end{equation}
In this case the protocol studied in the previous section allows  one to  distill the
singlet $|\Phi^-\rangle$ at an arbitrary time  with probability $e^{-2 \gamma t}/2$. The protocol run as follow:
a) Divide the particle into two similar sets, one as source and the other as ancilla. b) Alice applies a NOT operation on the ancilla c) Both observers apply a C-NOT operation regarding the source as control and the ancilla as target. b) Each observer measures the ancilla waiting for the result $+-$. c) If the result is successful the source is left in the state $|\Phi^+\rangle$  d) Alice applies a $S_z$ Pauli rotation in order to obtain the singlet $|\Phi^-\rangle$  as final state.

\section{The singlet in thermal baths.}

The master equation for a system composed  of two two-level particles in the presence of two independent thermal baths is
\begin{eqnarray}
\dot{\rho} &= &\frac{\gamma}{2}\left( (n+1) ( 2 \sigma_a \rho \sigma_a^{\dagger} -\sigma_a^{\dagger}\sigma_a \rho -\rho \sigma_a^{\dagger}\sigma_a )+ n ( 2 \sigma_a^{\dagger}  \rho \sigma_a -\sigma_a \sigma_a^{\dagger}\rho -\rho \sigma_a  \sigma_a^{\dagger}) \right.\nonumber\\
&+&\left. (n+1) ( 2 \sigma_b \rho \sigma_b^{\dagger} -\sigma_b^{\dagger}\sigma_b \rho -\rho \sigma_b^{\dagger}\sigma_b )+ n ( 2 \sigma_b^{\dagger}  \rho \sigma_b-\sigma_b \sigma_b^{\dagger}\rho -\rho \sigma_b  \sigma_b^{\dagger}) \right)
\end{eqnarray}
where we are assuming  that both baths have the same temperature so they have the same average  number $n$ of  thermal photons.
The solution of this equation with a singlet as initial condition is 
\begin{equation}
\rho=\left(\begin{array}{cccc}
\frac{(1+c)}{4}+\frac{d}{2} & 0 & 0 & 0\\
0 & \frac{(1-c)}{4}&\frac{a}{2}& 0\\
0 & \frac{a}{2} & \frac{(1-c)}{4} & 0\\
0 & 0 & 0 & \frac{(1+c)}{4}-\frac{d}{2}
 \end{array}\right)
\end{equation}
with
\begin{equation}
d(t) =  \frac{e^{-\gamma (1+2n) t}-1}{(1+2n)}
\end{equation}
\begin{equation}
a(t)= - e^{-\gamma (1+2n) t}
\end{equation}

\begin{equation}
 c(t) =
\frac{e^{-2 (2 n+1) t} \left(e^{2 (2 n+1) t}-2 e^{2 n t+t}-4 n (n+1)\right)}{(2 n+1)^2}
\end{equation}
The concurrence of this state is given
by:
\begin{equation}
C(t) = {\it max} \{ 0, C_{1}(t)\}
\end{equation}
 with 
\begin{equation}
C_1 (t) =  -a(t)-\frac{1}{4} \sqrt{(1+c(t))^2-4 (d(t))^2)}
\end{equation}
If one now applies exactly the same protocol described in the previous section the source is left in the following distilled state
\begin{equation}
\rho_d=\frac{1}{P(t)}\left(\begin{array}{cccc}
P_1(t) P_2(t) & 0 & 0 & 0\\
0 & (P_3+P_4)^2 & \frac{-(P_3-P_4)^2}{4}& 0\\
0 & \frac{-(P_3-P_4)^2}{4} & (P_3+P_4)^2 & 0\\
0 & 0 & 0 & P_1(t) P_2(t)
 \end{array}\right)
\end{equation}
where  $P(t) = 2 P_1(t) P_2(t)+\frac{(P_3-P_4)^2}{2}$ is the success probability and 
\begin{equation}
P_1 =  \frac{(1+c)}{4}+\frac{d}{2}
\end{equation}
\begin{equation}
P_2 =  \frac{(1+c)}{4}-\frac{d}{2}
\end{equation}
\begin{equation}
 P_3=  \frac{1}{2}\frac{(1-c)}{4}+\frac{a}{2}
\end{equation}\
\begin{equation}
 P_4=  \frac{1}{2}\frac{(1-c)}{4}-\frac{a}{2}
\end{equation}
are the eigenvalues of $\rho$.
The concurrence of the distilled state is given by
\begin{equation}
C_d(t) = {\it max} \{ 0, C_{2}(t)\}
\end{equation}
 with 
\begin{equation}
C_2 (t) =  \frac{(P_3-P_4)^2}{2 P}-\frac{P_1 P_2}{P} 
\end{equation}
In the Figure (3) we plot both the original and the distilled concurrence. As one can see the gain in entanglement is appreciable.
\begin{figure}\label{aaa}
\centerline{\includegraphics[scale=1.0]{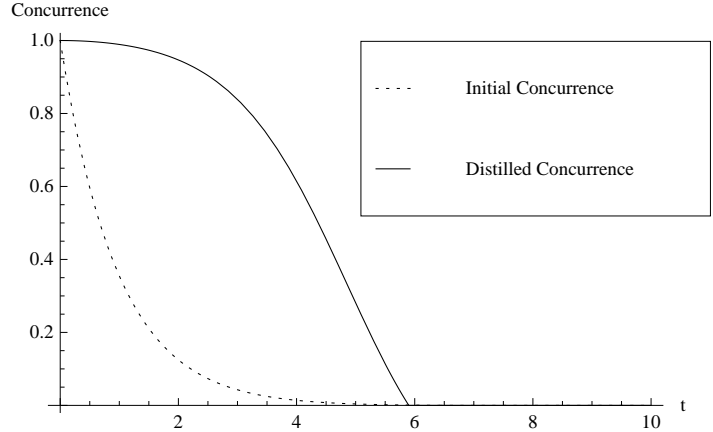}}
\caption{Non distilled Concurrence and distilled concurrence for initial state the singlet and thermal bath N=0.001}
\end{figure}

\section{Conclusions}

We have presented a classification of quasi-separable states in $2\times 2$ systems that identifies the set of non quasi-separable states with part of the non diagonal mixed states class. More precisely: a non quasiseparable state is given by setting the four parameters of the non diagonal state equal to $b=c=d=\frac{a}{2}$. Having characterized non quasiseparable states, we presented a distillation protocol which allows to increase the entanglement of the original state using a finite number of copies. For systems in vacuum it is possible to recover a maximally entangled state with probability greather than zero but smaller than $\frac{1}{2}$. By changing the accepted results in the measure of the ancilla, it is possible to obtain a final state with increased entanglement (but not maximally entangled) and probability greater than $\frac{1}{2}$. Finally, this protocol can also be applied in the case of non zero temperature. In this situation the system evolves with a comparatively sustained entanglement before the entanglement sudden dead time, although the probability rapidly falls bellow one half.

\section{Acknowledgments}
We wish to thank J. Stephany for his constructive remarks and precious help. This work was supported by Did-Usb Grant Gid-30 and by Fonacit Grant No G-2001000712.

\end{document}